\documentclass[useAMS,amssymb,usenatbib]{mn2e}
\usepackage{epsfig}
\usepackage{color}
\usepackage{graphicx}
\usepackage{tabularx}

\newcommand{\Ms}{\mathrm{M}_\odot}

\title[]{What is the (Dark) Matter with Dwarf Galaxies?}

\author[Sawala et al.] { Till Sawala$^{1}$\thanks{E-Mail:
    till@mpa-garching.mpg.de}, Qi Guo$^{3}$, Cecilia
  Scannapieco$^{2}$, Adrian Jenkins$^{3}$ and Simon White$^{1}$
  \\ $^{1}$Max-Planck Institute for Astrophysics,
  Karl-Schwarzschild-Strasse 1, 85748 Garching, Germany
  \\ $^{2}$Astrophysikalisches Institut Potsdam, An der Sternwarte 16,
  14482 Potsdam, Germany \\ $^{3}$Institute for Computational
  Cosmology, Department of Physics, University of Durham, South Road,
  Durham DH1 3LE, UK}

\begin{document}

\date{Accepted 2010 December 7. Received 2010 December 7; in original
  form 2010 March 1}

\pagerange{\pageref{firstpage}--\pageref{lastpage}} \pubyear{2010}

\maketitle

\label{firstpage}

\begin{abstract}
We present cosmological hydrodynamical simulations of the formation of
dwarf galaxies in a representative sample of haloes extracted from the
Millennium-II Simulation. Our six haloes have a $z=0$ mass of
$\sim10^{10}\Ms$ and show different mass assembly histories which are
reflected in different star formation histories. We find final stellar
masses in the range $5\times10^7 - 10^8\Ms$, consistent with other
published simulations of galaxy formation in similar mass haloes. Our
final objects have structures and stellar populations consistent with
observed dwarf galaxies. However, in a $\Lambda$CDM universe, $10^{10}
\Ms$ haloes must typically contain galaxies with much lower stellar
mass than our simulated objects if they are to match observed galaxy
abundances. The dwarf galaxies formed in our own and all other current
hydrodynamical simulations are more than an order of magnitude more
luminous than expected for haloes of this mass. We discuss the
significance and possible implications of this result.
\end{abstract}

\begin{keywords}
cosmology: theory -- galaxies: dwarf -- galaxies: formation --
galaxies: evolution -- galaxies: luminosity function, mass function --
methods: N-body simulations
\end{keywords}

\section{Introduction}
Dwarf galaxies are by far the most abundant type of galaxy in the
Local Group and in the Universe. They span a large range of stellar
masses, morphologies and star formation histories. The largest dwarf
irregulars such as the large Magellanic Cloud have stellar masses of
$\sim 10^9 \Ms$, rotationally supported and HI-rich disks, and strong
ongoing star formation. In contrast, dwarf spheroidal galaxies have
stellar masses from $10^7 \Ms$ to below $10^3 \Ms$, they possess no
interstellar gas, and they show no sign of rotational support or
ongoing star formation.

The number of dwarf galaxies observed in the Local Group continues to
grow as new, `ultra-faint' satellite galaxies are discovered
\citep[e.g.][]{Martin-2006, Chapman-2008, Belokurov-2010}. Estimates
using luminosity functions corrected for completeness and bias predict
the total number of faint satellites to be an order of magnitude
higher still \citep{Tollerud-2008, Koposov-2008}. Nevertheless, this
is still much smaller than the total number of dark matter subhaloes
found in high-resolution simulations of the standard $\Lambda$CDM
cosmology \citep[e.g.][]{Klypin-1999, Moore-1999, Diemand-2007,
  Springel-2008}. This difference has become known as the ``Missing
Satellites Problem''. It may only be an apparent discrepancy, however,
since it can be removed if one accounts for the fact that not all
low-mass subhaloes must contain stars, and those that do may have very
high mass-to-light ratios. Several astrophysical mechanisms have been
suggested that can lead to a number of visible satellite galaxies
similar to that observed. Perhaps haloes were able to form a few stars
initially, but the baryonic components of all haloes below some
critical mass were subsequently destroyed by supernova feedback
\citep[e.g.][]{Larson-1974, Dekel-1986, Ferrara-2000}. Alternatively
(or perhaps additionally) photoionisation may have prevented star
formation in the smallest haloes \citep[e.g.][]{Efstathiou-1992,
  Somerville-2002, Hoeft-2006, Simon-2007}. As \cite{Sawala-2010} have
shown, these two mechanisms can combine to produce very high
mass-to-light ratios in haloes of $10^9 \Ms$ and below, perhaps
reconciling the number of very faint dwarf galaxies produced in
$\Lambda$CDM simulations with the observations.

In this work, we turn our focus to more massive dwarf galaxies, and
follow the evolution of the objects that form in haloes of
$10^{10}\Ms$.  Our initial conditions are based on six haloes selected
from the Millennium-II Simulation
\citep[MS-II,][]{Boylan-Kolchin-2009}, and resimulated at high
resolution using smoothed particle hydrodynamics (SPH). Our
simulations include cooling and star formation, supernova feedback,
metal-enrichment and a cosmic UV background. Starting at redshift
$z=49$, we are able to follow the formation of each individual halo
and its central galaxy in their full cosmological context, all the way
to $z=0$.

On the other hand, the large volume of our parent simulation allows us
to verify that our sample of resimulated haloes is representative of
haloes of similar mass, and to predict a stellar mass~-- halo mass
relation that can be tested against observation. With a box size of
137~Mpc and a mass resolution of $9.4\times10^6\Ms$, the MS-II has
sufficient dynamic range to capture the statistics of the assembly of
dark matter haloes between $10^9$ and $10^{14}\Ms$. By comparing its
halo/subhalo mass function to the observed SDSS stellar mass function
of \cite{Li-2009}, \cite{Guo-2010} derived a typical mass-to-light
ratio for each halo mass. This analysis assumes a monotonic
relationship between halo mass and galaxy mass with relatively small
scatter, but does not rely on any other assumptions about the
processes involved in galaxy formation. We use its result to test the
viability of our simulations and the underlying physical model as a
description of the formation of ``typical'' $\Lambda$CDM dwarf
galaxies.

The present work constitutes the first direct comparison of high
resolution, hydrodynamical simulations of individual dwarf galaxies
with the observed abundance of such objects. We combine the ability to
follow star formation self-consistently in individual objects with the
ability to draw conclusions about the general population of dwarf
galaxies.\\

This paper is organised as follows: We begin in
Section~\ref{sec:previous} by reviewing the current status of
simulations of the formation of dwarf galaxies. Section~\ref{sec:ICs}
describes the selection of haloes for resimulation and the generation
of our high resolution initial conditions, while the numerical methods
of our hydrodynamic simulations are discussed briefly in
Section~\ref{sec:methods}. In Section~\ref{sec:evolution}, we show
results for six haloes of final mass $10^{10}\Ms$, and compare the
properties of the galaxies to previous work, and to observation. In
Section~\ref{sec:sams}, we consider the predictions of our simulations
for the stellar mass~-- halo mass relation and discuss the discrepancy
with that inferred from comparing the observed stellar mass function
to the halo abundance in $\Lambda$CDM simulations. We conclude with a
summary and interpretation of our results in
Section~\ref{sec:summary}.

Unless stated otherwise, where we refer to the mass of a {\it galaxy},
we mean the stellar mass M$_\star$, whereas the mass of a {\it halo}
includes the total dynamical mass enclosed within r$_{200}$, the
radius that defines a spherical overdensity 200 times the critical
density of the universe. When quoting the results for our own
simulations, we always use physical mass units of $\Ms$, assuming
h~$=0.73$.

\begin{figure*}
  \vspace{-.2in}
  \begin{center}
    \begin{tabular}{lcccc}
       \hspace{-.2in} \includegraphics*[scale = .415]{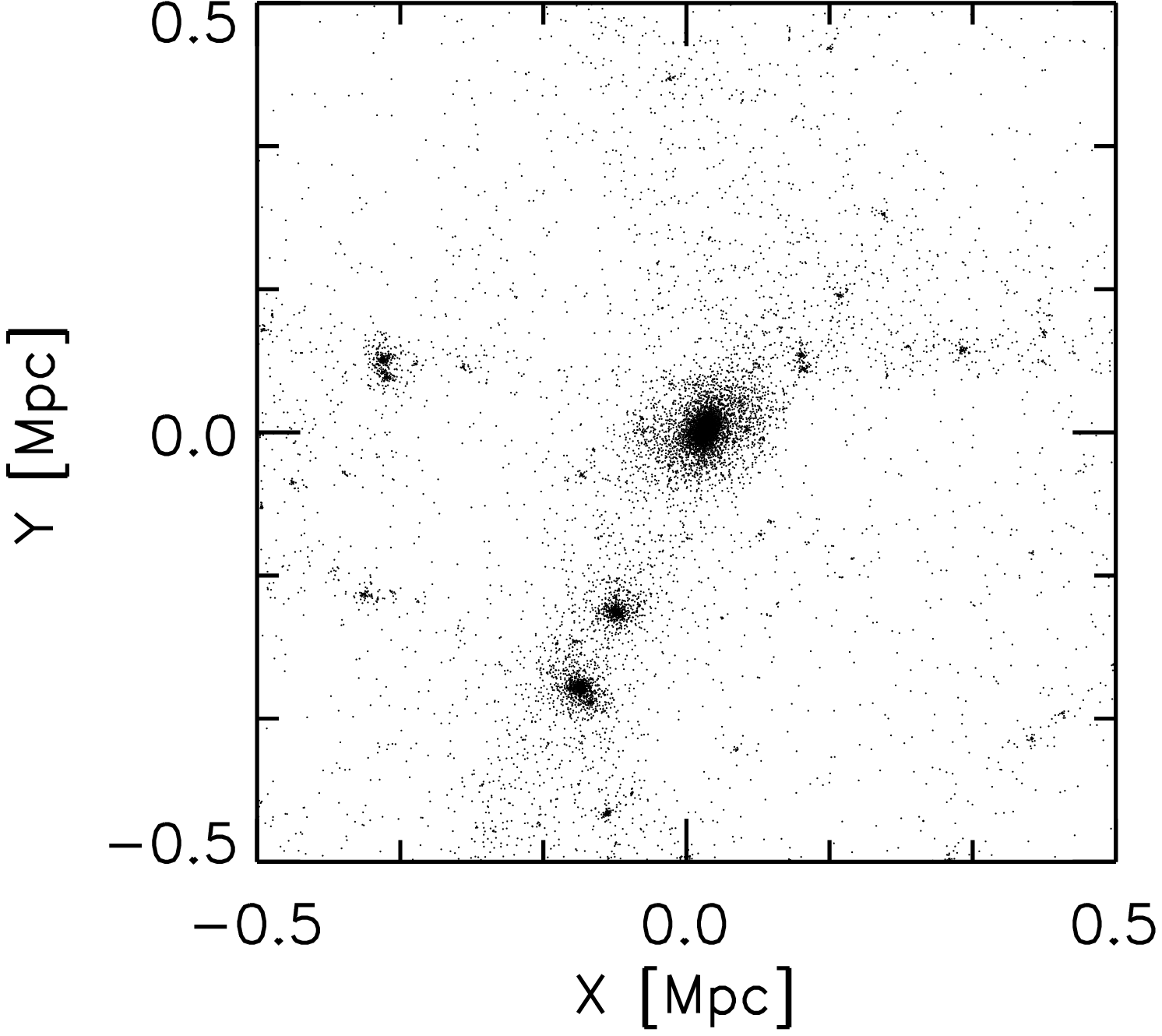} &
       \hspace{-.5in} \includegraphics*[scale = .415]{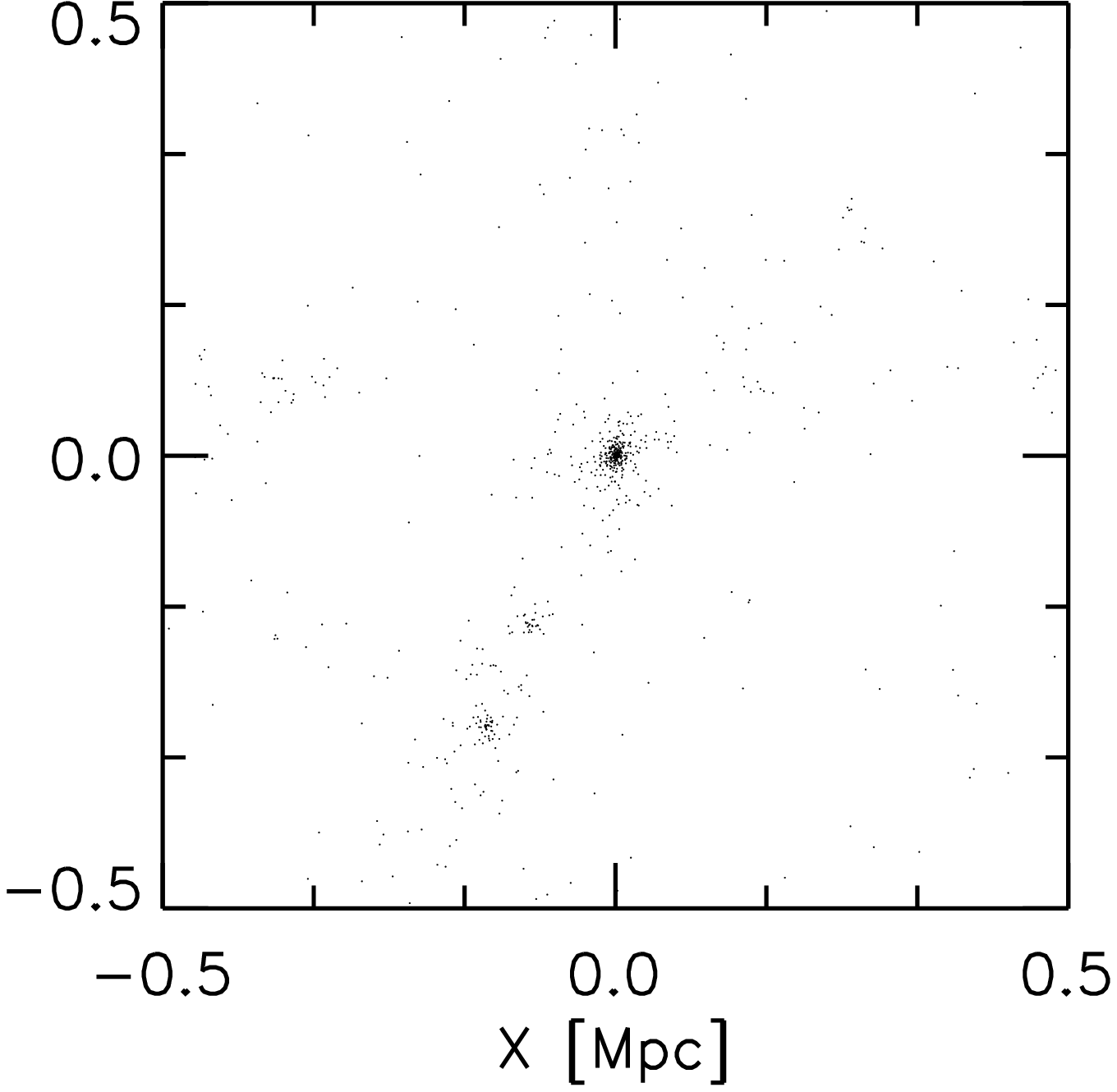} &
       \hspace{-.5in} \includegraphics*[scale = .415]{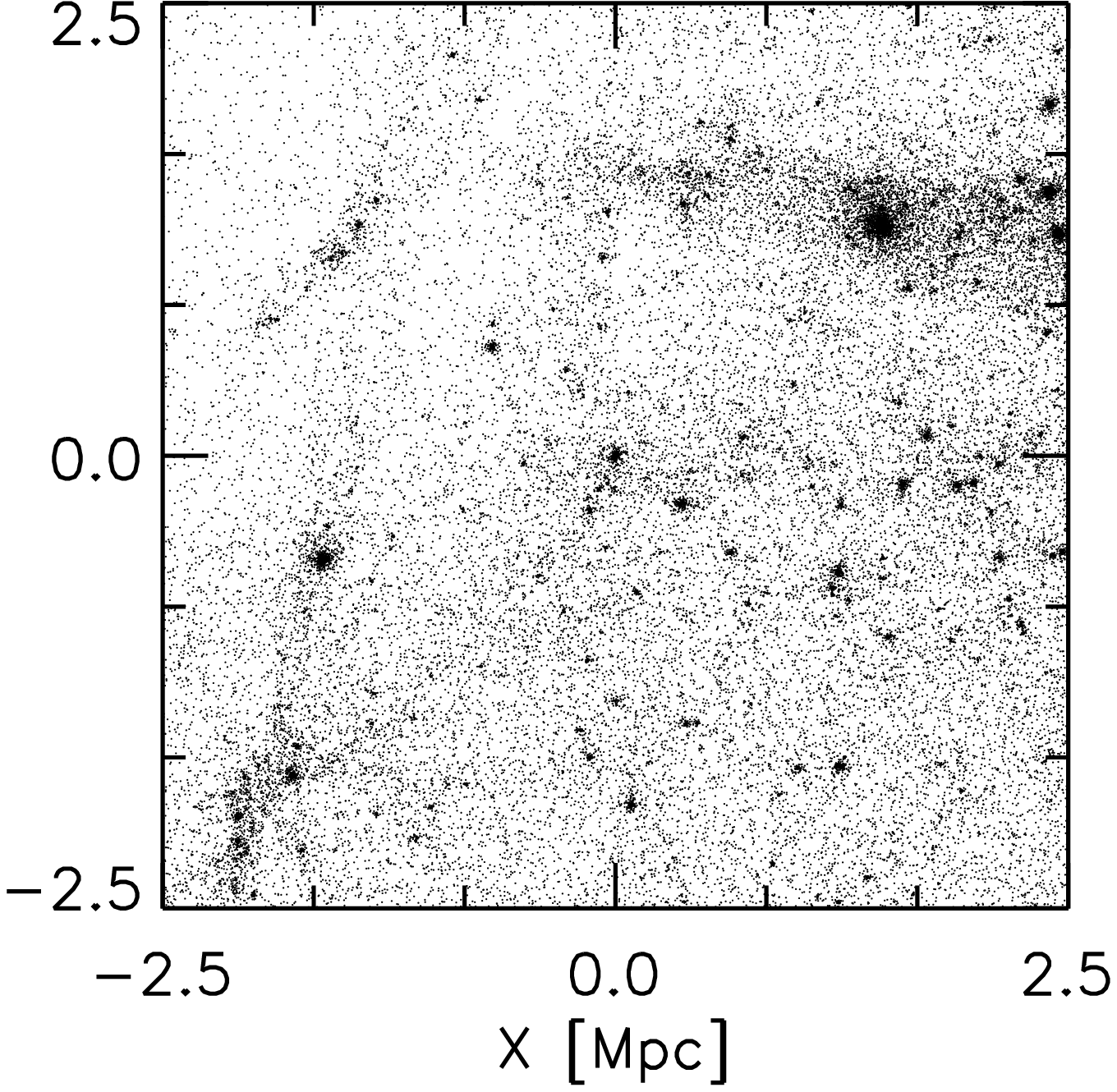} 
    \end{tabular}
  \end{center}
  \vspace{-.3in}
  \caption{Comparison of Halo~4 at $z=0$ in a pure dark matter
    resimulation and in the parent Millennium-II Simulation. The left
    panel shows the position of $0.5\%$ of the particles in a box of
    sidelength 1~Mpc in the resimulation, while the central panel
    shows the position of all particles within the same region in the
    MS-II. The panel on the right shows all particles in a box of
    5~Mpc in the MS-II, with Halo~4 in the centre. All three panels
    are centred on the same absolute coordinates for the parent box
    of sidelength 137~Mpc, showing the position of the halo to be in
    perfect agreement. The FoF mass of the halo also agrees to within
    less than $1\%$. A comparison of the left and the central panel
    reveals the additional substructure resolved in the
    resimulation.\label{fig:compare_m2}}
\end{figure*}

\section{Review of previous work}\label{sec:previous}

\begin{table}
\caption{Results of earlier numerical simulations\label{table:other}}
\begin{tabularx}{\columnwidth}{lrrX}
  \hline Reference & M$_\star$ & M$_\mathrm{tot}$ & $v_c$ or $\sigma_\star$ \\ 
   &\scriptsize{$[10^7\Ms]$} &\scriptsize{$[10^9\Ms]$}  &\scriptsize{[km s$^{-1}$]}\\
  \hline  \vspace{-.05in}\\
  Pelupessy et al. (2004)\footnotemark[1] & $18$ & $15$ & 80 \\
  Stinson et al. (2007)\footnotemark[2] & $7.86$ & $5.0$ & 15.1 \\
  Stinson et al. (2007)\footnotemark[2] & $22$ & $8.6$ & 20.1 \\
  Stinson et al. (2007)\footnotemark[2] & $38.6$ & $14$ & 29.9 \\
  Stinson et al. (2009)\footnotemark[3] & $1.72$ & $14$ & 16.8 \\
  Valcke et al. (2008)\footnotemark[4] & $57.9$ & $4.1$ & 35.2 \\
  Valcke et al. (2008)\footnotemark[4] & $48.8$ & $4.1$ & 30.9 \\
  Mashchenko et al. (2008)\footnotemark[5] & $1.0$ & $2.0$ & -- \\
  Governato et al. (2010)\footnotemark[6] & $48$ & $35$ & 56  \\
  Governato et al. (2010)\footnotemark[6] & $18$ & $20$ & 54  \\
  \hline\vspace{.1in}
\end{tabularx}

\small{Notes: Col.~2: Stellar mass, Col.~3: Halo mass (M$_{200}$),
  Col.~4: Maximum rotation velocity$^{1,6}$ or 1-D velocity
  dispersion$^{2,3,4}$. All quantities are measured at $z=0$, except
  for Mashchenko et al., where the halo mass is at $z=5$ and stellar
  mass at $z=6.2$.}

\small{Remarks: $^1$Static initial conditions set up to reproduce the
  dwarf irregular galaxy DDO~47; $^{2,3}$Static NFW profiles with
  initial baryon fractions of $10\%^2$ and $1\%^3$; $^4$Runs DH01 and
  DH02, assuming static, simplified Kuz'min Kutuzov profiles with
  $a=c=4$~kpc (DH01) and 6~kpc (DH02).}
\end{table}

Earlier examples of numerical studies of dwarf galaxy formation and
evolution in $\sim10^{10}\Ms$ haloes include simulations by
\cite{Pelupessy-2004}, \cite{Stinson-2007, Stinson-2009},
\cite{Valcke-2008}, \cite{Mashchenko-2008} and \cite{Governato-2010}.

The first three have investigated the evolution of dwarf galaxies
embedded in dark matter haloes of constant
mass. \citeauthor{Pelupessy-2004} used initial conditions modelled
after dwarf irregular galaxy DDO 47, set up with a stellar disk of
$1.8\times10^8\Ms$ and a gas disk of $1.9\times10^8\Ms$ inside a dark
matter halo of $1.5\times10^{10}\Ms$. Thus, the stellar mass to halo
mass ratio is not a result of their simulation, but was chosen {\it a
  priori}. They show that the star formation behaviour in such a
system is consistent with observations. \citeauthor{Valcke-2008}
studied the formation of dwarf elliptical galaxies assuming cored
initial dark matter profiles following \cite{Dejonghe-1988}, and gas
at a mass fraction of $17.5\%$. Cooling, star formation and feedback
are included in their simulations, making their final stellar masses
of $4.9-5.8 \times10^8 \Ms$ for haloes of $4.1\times10^{9}\Ms$ a
direct prediction of their models. \citeauthor{Stinson-2007} also
assume fixed dark matter profiles in their initial conditions, and
perform simulations that include cooling, star formation and
feedback. In \cite{Stinson-2007}, a fixed initial baryon fraction of
$10\%$ is assumed, and stellar masses of
$7.9\times10^7-3.9\times10^8\Ms$ are produced in haloes of
$5\times10^9-1.4\times10^{10}\Ms$. In \cite{Stinson-2009}, the baryon
fraction is varied, and we also include their result with a very low
initial baryon fraction of $1\%$, that leads to a smaller stellar
mass.\citeauthor{Valcke-2008} and \citeauthor{Stinson-2009} do not
include a UV background, which may contribute to the high star
formation efficiency in their simulations.

\cite{Mashchenko-2008} and \cite{Governato-2010} both performed
simulations that include the formation of the dark matter halo in a
cosmological volume. \citeauthor{Mashchenko-2008} used constrained
initial conditions, aimed at reaching a halo mass of $10^9\Ms$ at
$z=6$, and followed the evolution up to $z=5$. At this time, their
halo reached a mass of $2\times10^9\Ms$, with $10^7\Ms$ of stars
formed. By comparison with the typical evolution of haloes in the
MS-II (see Figure~\ref{fig:haloes}), we note that this is consistent
with a mass of $10^{10}\Ms$ at $z=0$. The extent of additional star
formation up to $z=0$ is unknown, however. Most recently,
\citeauthor{Governato-2010} have performed hydrodynamical simulations
of two dwarf irregular galaxies at very high resolution, which they
follow up to $z=0$. This makes these most comparable to our own
simulations, and also makes their results most directly comparable to
observed, present-day dwarf galaxies. Their simulations start with
values of $\Omega_\mathrm{m}=0.24$, $\Omega_\mathrm{b}=0.042$, and
predict stellar masses of $1.8$ and $4.8\times10^8\Ms$ in two haloes
of $2.0$ and $3.5\times10^{10}\Ms$, respectively.

While these five sets of simulations vary in the setup of the initial
conditions, the cooling, star formation and feedback recipes, the
treatment of the cosmic UV background, the simulation code and the
numerical resolution, they all predict final stellar masses consistent
with $\sim10^8\Ms$ for dark matter haloes of $\sim10^{10}\Ms$. We give
an overview of some of the relevant properties of these simulations in
Table~\ref{table:other}.

\section{Initial Conditions}\label{sec:ICs}
The parent simulation, as well as our high-resolution resimulations,
are performed in the context of a $\Lambda$CDM cosmology, with
$\Omega_\Lambda=0.75$, $\Omega_\mathrm{m}=0.25$, h$=0.73$ and
$\sigma_8=0.9$, identical to the values used for the original
Millennium Simulation \citep{Springel-Millennium}.

The Millennium-II Simulation followed structure formation in a volume
of $137^3$~Mpc$^3$ using $2160^3$ dark matter particles and periodic
boundary conditions. This corresponds to a mass resolution of
$9.43\times10^6 \Ms$, and a force resolution of 1.37~kpc. At $z=0$, it
contains about 12 million friends-of-friends (FoF) haloes with at
least 20 particles, corresponding to a minimum resolved halo mass of
$1.9\times10^8\Ms$. Haloes of $\sim10^{10}\Ms$, the mass at $z=0$ that
we select for our resimulations, are resolved with over $10^3$
particles.

\begin{figure}
  \vspace{-.2in}
  \hspace{-.2in} \includegraphics[scale = .462]{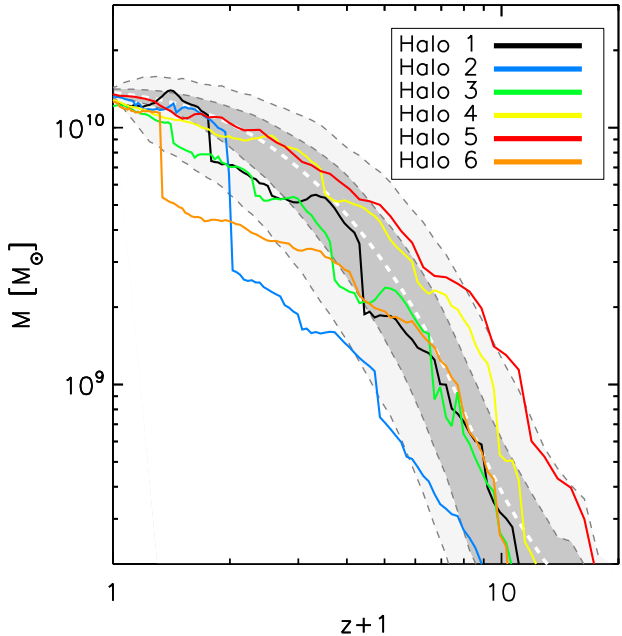}\\
  \vspace{-.2in}
  \caption{Evolution of FoF-halo mass as a function of redshift in our
    pure dark matter resimulations. The solid coloured lines show the
    mass accretion history of the six haloes we have re\-simulated at
    high resolution. Overplotted as a thick dashed line is the mean
    halo mass from the Millennium-II Simulation, for all haloes of
    similar final masses. Also shown are the $1\sigma$ and $2\sigma$
    upper and lower bounds at each redshift, in dark and light shades,
    respectively. It can be seen that there is a variety of assembly
    histories, both in the parent simulation, and in the sample of
    resimulated haloes. Haloes~3 and 6 have undergone recent major
    mergers, while Halo~5 formed significantly earlier than the five
    others. The variety within our sample is somewhat greater than
    expected for a random sub-sample of the MS-II, but there is no
    systematic bias in formation history. Formation redshifts, defined
    as the time when a halo reaches half its peak mass, lie between 1
    and 2.5, consistent with a median formation redshift of 2. Note
    that the halo mass m$_{200}$ can differ from the FoF-mass by
    $\sim5-15\%$, and due to outflows, the total mass in each of the
    six haloes is reduced to $\sim 10^{10}\Ms$ in the resimulations
    with gas.\label{fig:haloes}}
\end{figure}

Out of more than $10^4$ haloes within our mass range in the MS-II, we
identified 25 haloes as resimulation candidates, based on the
condition that all particles within twice the virial radius at
redshift $z=0$ were in a connected region and inside a sphere of
radius 0.67~Mpc in the initial conditions. Out of these, six haloes
were selected in order to study a varied but representative sample of
mass accretion histories (see Figure~\ref{fig:haloes}). The initial
conditions for the resimulations were generated by re-sampling the
region of interest with a high number of low mass dark matter
particles, while the remaining volume was sampled with increasingly
coarser resolution, sufficient to capture the long-range tidal
field. To account for the higher Nyquist frequency of the
resimulations, small-scale fluctuations were added to the displacement
and velocity fields of the original MS-II using the method of
second-order Lagrangian perturbation theory described in
\cite{Jenkins-2010}.

Figure~\ref{fig:compare_m2} shows the final distribution of dark
matter particles in slices centred on a dwarf halo in the MS-II and
in one of our high-resolution resimulations (pure dark matter). The
difference in mass between the central FoF halo in the resimulations
and in the MS-II is $< 1\%$, equivalent to $\sim10$ particles in the
parent simulation. The position and velocity are well reproduced, and
the agreement also extends to substructures outside the main halo. The
structure resolved in the MS-II is again found at the correct mass and
location, while some additional substructure is visible only in the
resimulation.

We have performed resimulations both with pure dark matter and with
gas particles added, splitting each high-resolution dark matter
particle at a mass ratio of $\Omega_\mathrm{b}=0.046$ to
$\Omega_\mathrm{DM}=0.204$. All resimulations start at $z=49$ and are
evolved up to $z=0$.  The hydrodynamical simulations include $1.1
\times 10^6$ high resolution dark matter particles of $8 \times10^4
\Ms$, and an equal number of gas particles of $1.8 \times10^4
\Ms$. The initial stellar particle mass is $9\times10^3\Ms$. At $z=0$,
the individual haloes are resolved with more than $10^5$ particles
within r$_{200}$. The volume outside of the high-resolution region is
sampled with an additional $7.6 \times10^5$ dark matter particles of
varying mass to include the evolution of structure on large scales.

To check for possible biases due to our selection method, we have
compared our candidates for resimulation to the total population of
similar mass haloes in the MS-II. Figure~\ref{fig:haloes} shows as
solid lines the merger histories of the six selected dark matter
haloes in the high resolution simulations, together with the typical
mass accretion history, derived from merger trees of $\sim10^4$
randomly selected, similar mass haloes. At each redshift, the inner
and outer grey regions indicate the 3rd, 16th, 84th and 97th
percentiles, equivalent to $1\sigma$ and $2\sigma$ deviations from the
mean mass for a Gaussian mass distribution. It can be seen that the
variance within our sample is higher than the expected variance within
a random sub-sample of haloes. This allows us to follow the evolution
of haloes with a range of merger histories in a limited number of
simulations. However, there is no systematic bias in the mass
accretion history of our haloes, so our sample can be considered a
reasonably unbiased representation of $10^{10}\Ms$ haloes in the
MS-II.

\section{Numerical Methods}\label{sec:methods}
The high-resolution simulations presented here have been performed
using the Tree-PM code \textsc{GADGET-3} \citep{Springel-2005,
  Springel-2008}, which includes gravity and smoothed particle
hydrodynamics. As an extension, metal-dependent cooling, star
formation, chemical enrichment and energy injection from type II and
type Ia supernovae have been implemented in the multiphase gas model
of \cite{Scannapieco-2005, Scannapieco-2006}. This model has
previously been used to study the formation both of large disk
galaxies \citep{Scannapieco-2008, Scannapieco-2009}, and of dwarf
spheroidal galaxies \citep{Sawala-2010}. In
Sections~\ref{sec:methods:softening} to~\ref{sec:ism}, we briefly
explain the most important characteristics of this code, and refer the
interested reader to the above references for a more detailed
description.

\subsection{Gravitational Softening}\label{sec:methods:softening}
In order to reduce two-body interactions arising from the particle
representation of the matter distribution, the gravitational potential
is modified by replacing the divergent $1/r^{2}$ dependence with
$1/(r^2 +\epsilon^2)$, where $\epsilon$ is the gravitational softening
scale \citep{Aarseth-1963}. The choice of $\epsilon$ represents a
compromise between the errors due to residual two-body effects, and
the loss of spatial resolution below several softening scales. We
begin our simulations with a softening length fixed in comoving
coordinates to $1/10$th of the mean interparticle spacing for each
particle type, corresponding to $\sim1$h$^{-1}$~kpc in the high
resolution region. After the collapse of the halo, we keep the
softening scale in this region constant in physical coordinates from
$z=7$, at a value of 155~pc. \cite{Power-2003} give a lower limit
$\epsilon_{acc} = r_{200}/\sqrt{N_{200}}$ to prevent strong
discreteness effects in haloes, which corresponds to $\sim140$~pc for
a $10^{10}\Ms$ object resolved with $N_{200}\sim 10^5$ particles. We
also resimulated one of our haloes, Halo~4, with a physical softening
scale of 77.5~pc, and checked that this did not alter the results
significantly.

\subsection{Cooling and UV Background}\label{sec:methods:cooling}
Above the hydrogen ionisation temperature of $10^4$~K, our gas cooling
model is based on metal-dependent cooling functions of
\cite{Sutherland-1993}. The model assumes collisional excitation
equilibrium, and does not include metal or molecular cooling below
$10^4$K. In addition, we include Compton cooling, which is the main
coolant at high redshift. It depends on the free electron density, as
well as on the temperature difference between the gas and the evolving
CMB.  For this purpose, the ionisation states of H, He, and the free
electron number density are computed analytically, following the model
of \cite{Katz-1996}. We have included UV background radiation in our
model, which adds a heating term to the net cooling function of the
partially ionised gas. In all simulations, the UV background is
present from $z=6$, and its spectral energy distribution and the time
evolution of its intensity follow the model of \cite{Haardt-1996}. A
test simulation of Halo~4 without the UV background produced over
twice as many stars by $z=1$, compared to the simulation which
includes UV radiation.

\subsection{Star Formation Criteria} \label{sec:sf}
Cold gas particles can spawn, or be converted into, star particles,
subject to certain conditions. We require the gas particle to be in a
region of convergent flow. In addition, we impose a physical density
threshold $\rho_c$ on the local gas density. The existence of a
threshold for star formation is motivated by observations
\citep[e.g.][]{Kennicutt-1989, Kennicutt-1998}. Calculations by
\cite{Quirk-1972} as well as numerical simulations, e.g. by
\cite{Katz-1996, Springel-2003, Bush-2008} and others have shown that
the observed Kennicutt-Schmidt relation can be reproduced in disk
galaxies by imposing a volume density threshold, even though different
values are assumed. \cite{Koyama-2009} demonstrated with
high-resolution simulations of the turbulent interstellar medium that
the star formation rate depends only weakly on the choice of $\rho_c$,
and values in the range 0.1~cm$^{-3}$ \citep{Stinson-2009} to
100~cm$^{-3}$ \citep{Governato-2010} can be found in the recent
literature. \citeauthor{Governato-2010} reported better convergence in
their high-resolution simulation with a choice of 100 compared to
0.1. In this work, we adopt a value of 10~cm$^{-3}$. We have also
tested a density threshold of 0.1~cm$^{-3}$, more similar to our own
previous work. In this case, star formation starts at higher redshift
and is less bursty. For Halo~4, the final stellar mass increases by
$\sim36\%$ with a threshold of 0.1~cm$^{-3}$. This difference is less
than the variance in stellar mass between individual haloes, and does
not qualitatively affect the stellar mass-halo mass ratio. The limited
effect of $\rho_c$ results from the fact that star formation is mostly
self-regulating in our simulations. We also impose a threshold
$\rho_\mathrm{g} / \overline{\rho_\mathrm{g}} \ge 10^4 $ on the local
gas overdensity relative to the cosmic mean, which ensures that star
formation only takes place in virialized regions even at very high
redshift.

\subsection{Star Formation Efficiency} \label{sec:sf-eff}
Subject to the constraints described in Section~\ref{sec:sf}, the star
formation efficiency is regulated by a single efficiency parameter
$c_\star$, so that the star formation rate density is given by
$\dot{\rho}_\star = c_\star \rho_g t_\mathrm{dyn}^{-1}$, where
$t_\mathrm{dyn}$ is the local gas dynamical time. The creation of an
individual stellar particle of mass $m_\star$ from a gas particle of
mass $m_g$ during the time interval $\Delta t$ is stochastic, with
probability given by:

$$ p_\star= \frac{m_g}{m_\star} \left[1 - \exp\left(-~c_\star
  \frac{\Delta t}{t_\mathrm{dyn}} \right)\right] $$

In simulations with radiative transfer, \cite{Gnedin-2008} found that
dust acts as a catalyst for molecular cloud formation, suggesting that
star formation may be less efficient in the low metallicity
environment of dwarf galaxies. As our simulations cannot follow cloud
formation, we do not take this into account and assume a constant
$c_\star$. \cite{Ricotti-2002} showed that if star formation is
strongly self-regulating, the star formation rate is determined
primarily by the thermodynamic properties of the gas and depends only
very weakly on $c_\star$. This result was confirmed in our previous
work \citep{Sawala-2010}, and in all simulations presented here, we
adopt our earlier value of $c_\star=0.05$.

Each star particle is produced with a single stellar population, whose
metallicity is inherited from the parent gas particle. We assume a
Salpeter initial mass function \citep{Salpeter-1955}, and calculate
stellar luminosities using the stellar synthesis model of
\cite{BC-2003}.

\subsection{Multiphase Interstellar Medium and Feedback} \label{sec:ism}
For each star particle, we determine the rate as well as the yields of
supernovae type II and type Ia. Chemical yields are calculated
separately for the two types, following \cite{Woosley-1995} and
\cite{Thielemann-1993}, respectively. Supernovae type II are assumed
to be instantaneous, while supernovae type Ia follow a uniform delay
time distribution between 100~Myrs and 1 Gyr. We assume a constant
energy production of $7\times10^{50}$ ergs per supernova, which is
released into the interstellar medium (ISM) as thermal energy.

The multiphase scheme of \cite{Scannapieco-2006} allows an overlap of
diffuse and dense gaseous components. This preserves the multiphase
structure characteristic of the ISM, in which components with very
different temperatures and densities coexist. It also avoids the
overestimation of density in diffuse gas near high density regions
which can cause a serious underestimate of its cooling time. The
decoupling is achieved by considering as neighbours in the SPH
smoothing kernel only gas particles with similar thermodynamical
properties, as defined by the ratio of their entropic functions.

\begin{figure*}
  \begin{center}
     \begin{tabular}{ccc}
      \vspace{-.6in}
      \hspace{-.35in} \includegraphics*[scale = .275]{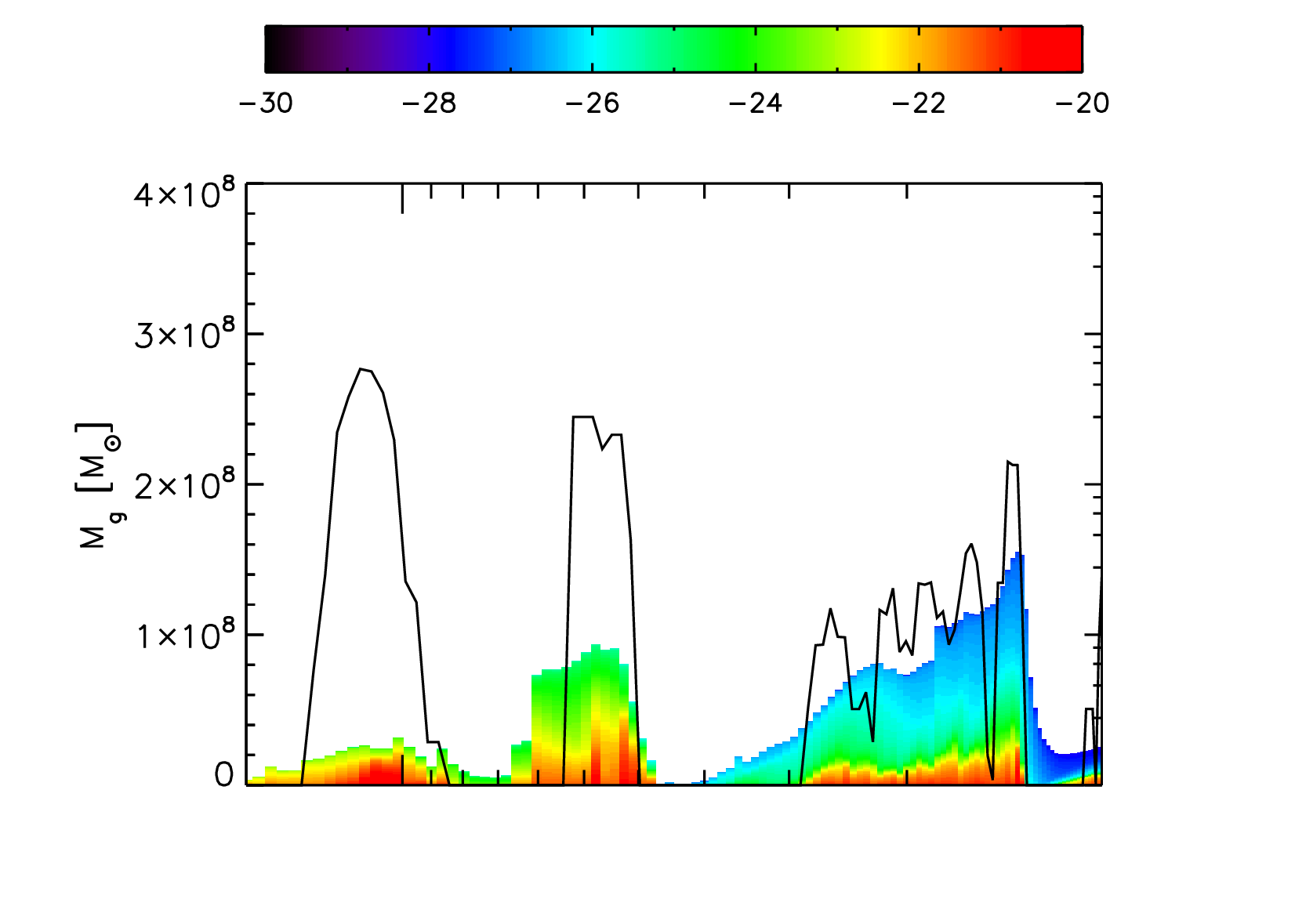} &
      \hspace{-1.15in} \includegraphics*[scale = .275]{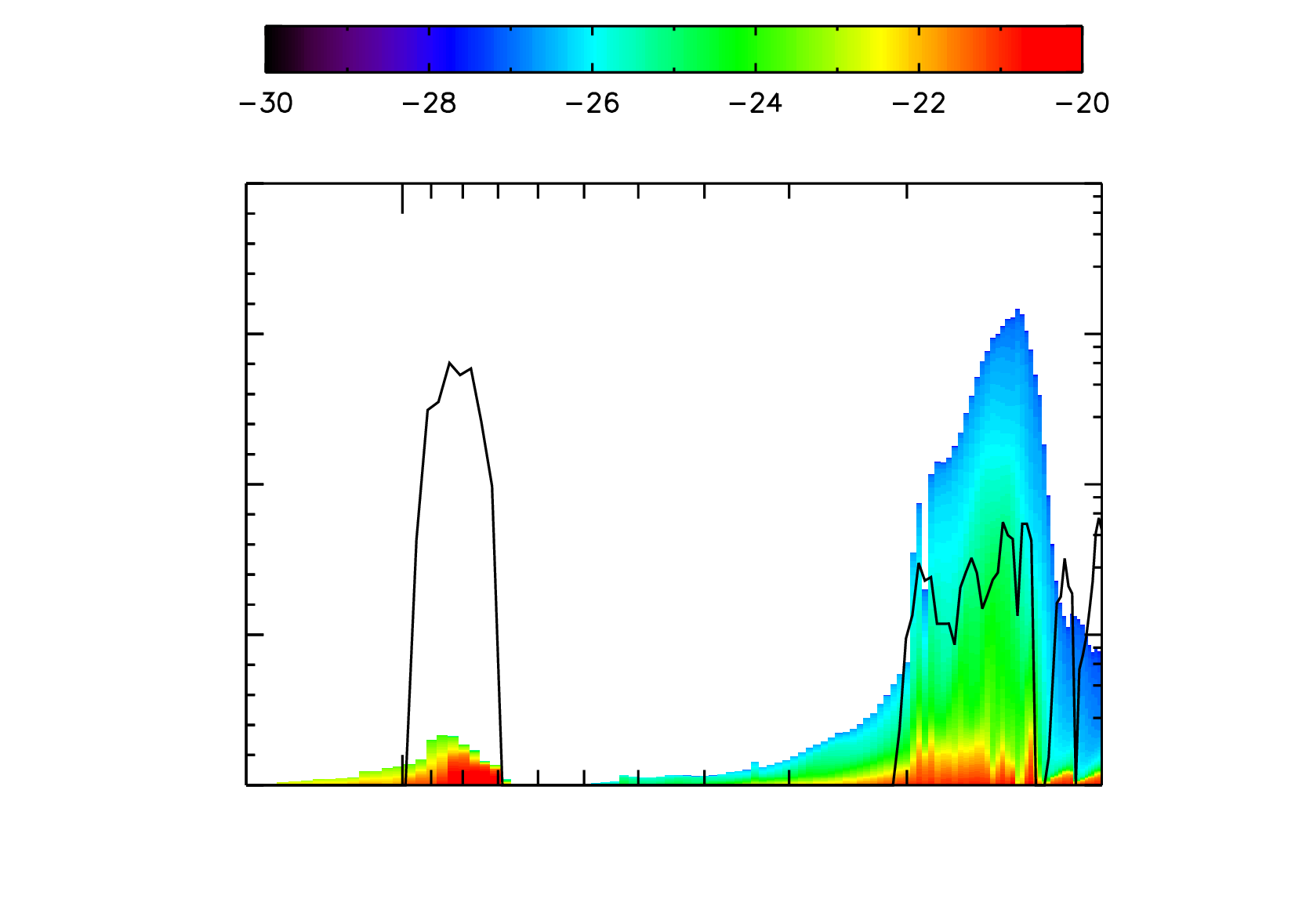} &
      \hspace{-1.15in} \includegraphics*[scale = .275]{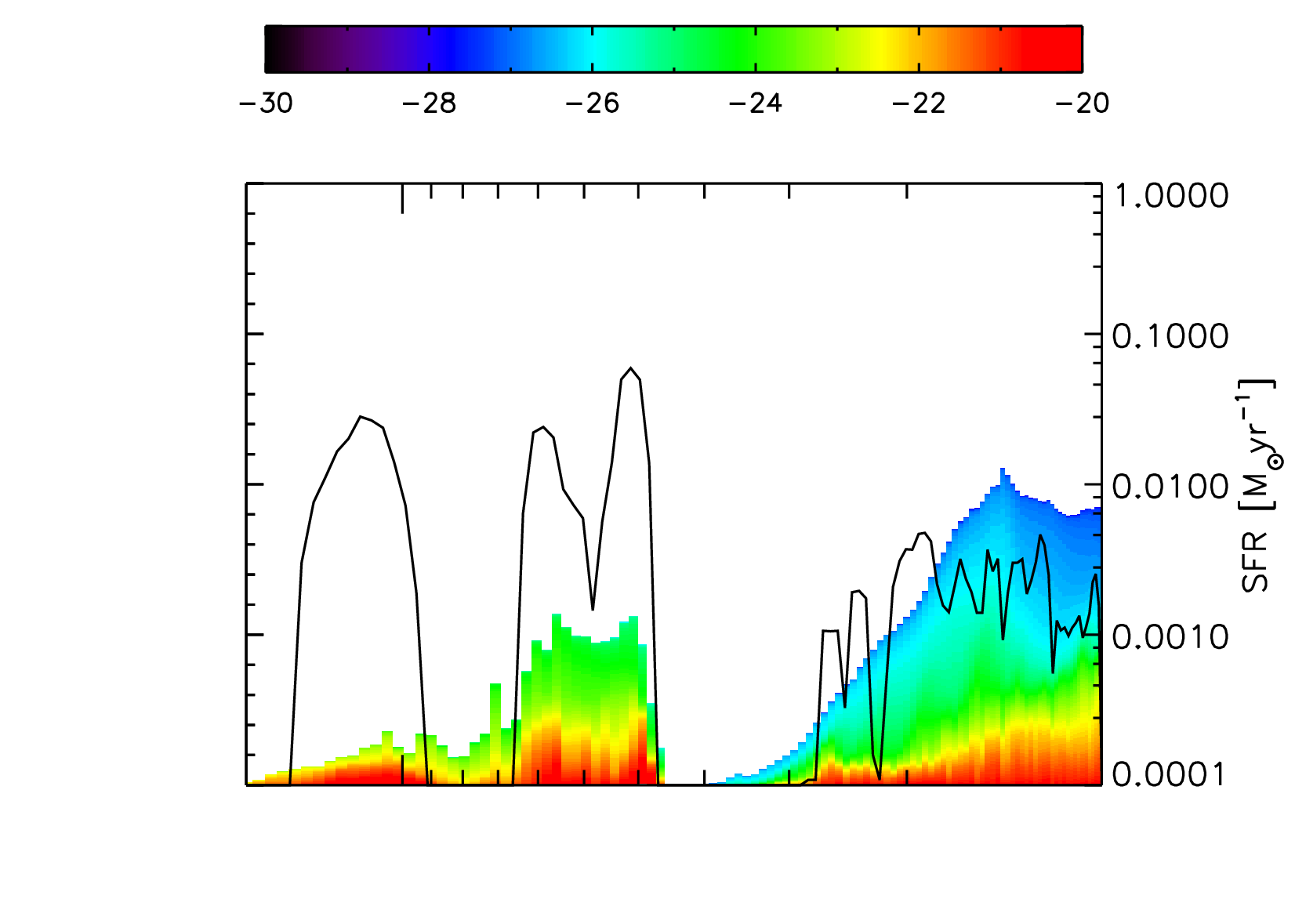} \vspace{-.1in} \\
      \hspace{-.35in} \includegraphics*[scale = .275]{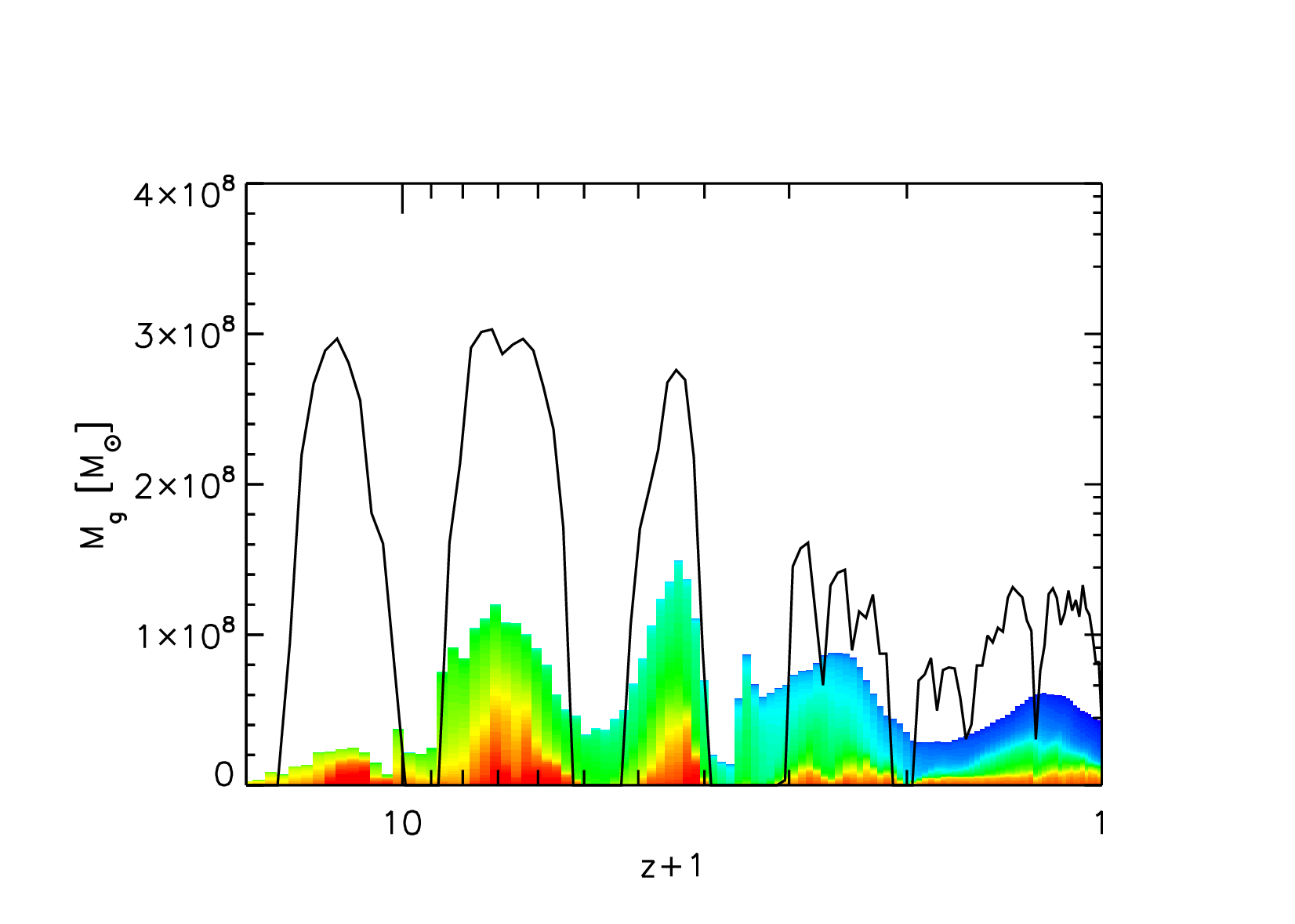} &
      \hspace{-1.15in} \includegraphics*[scale = .275]{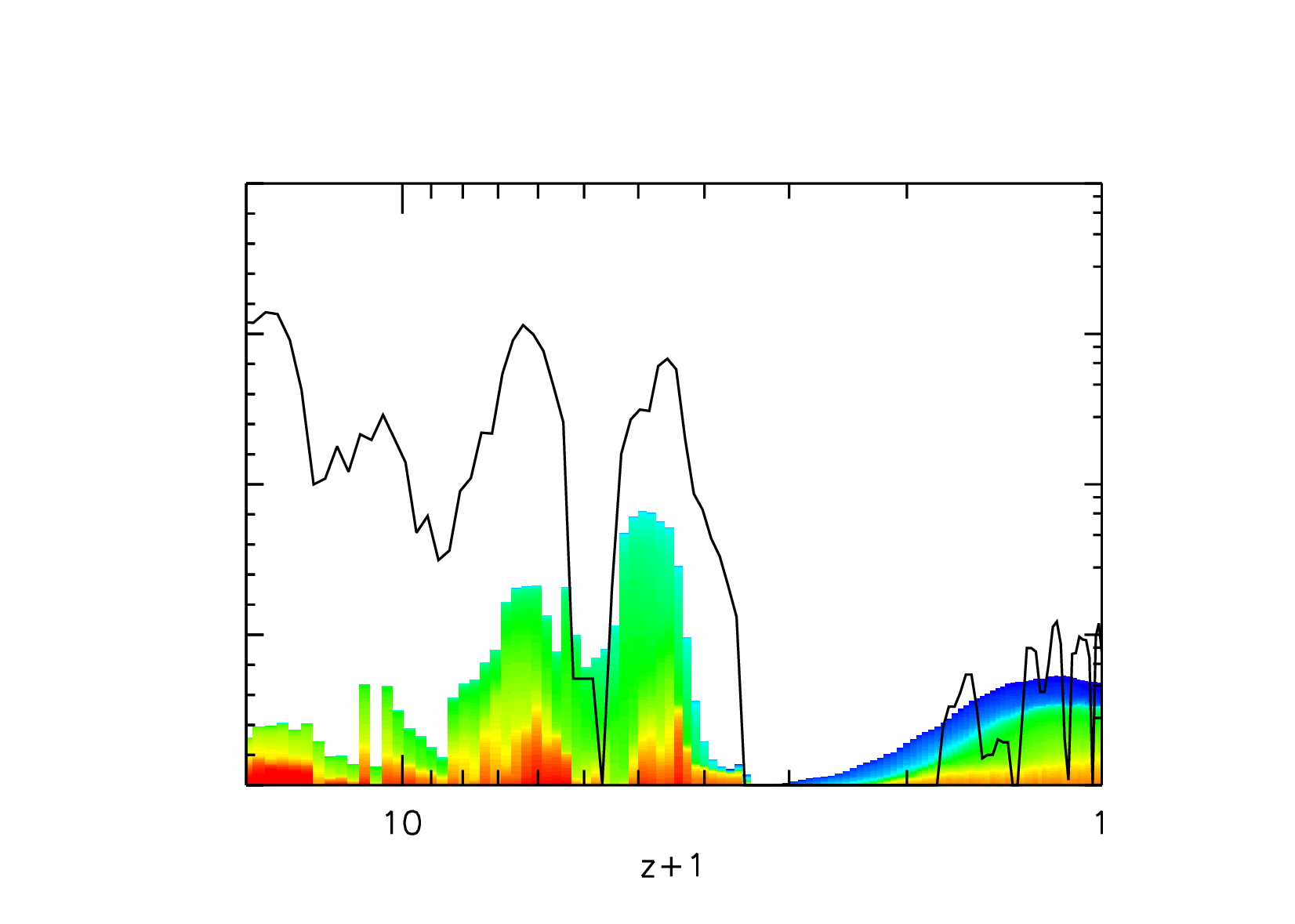} &
      \hspace{-1.15in} \includegraphics*[scale = .275]{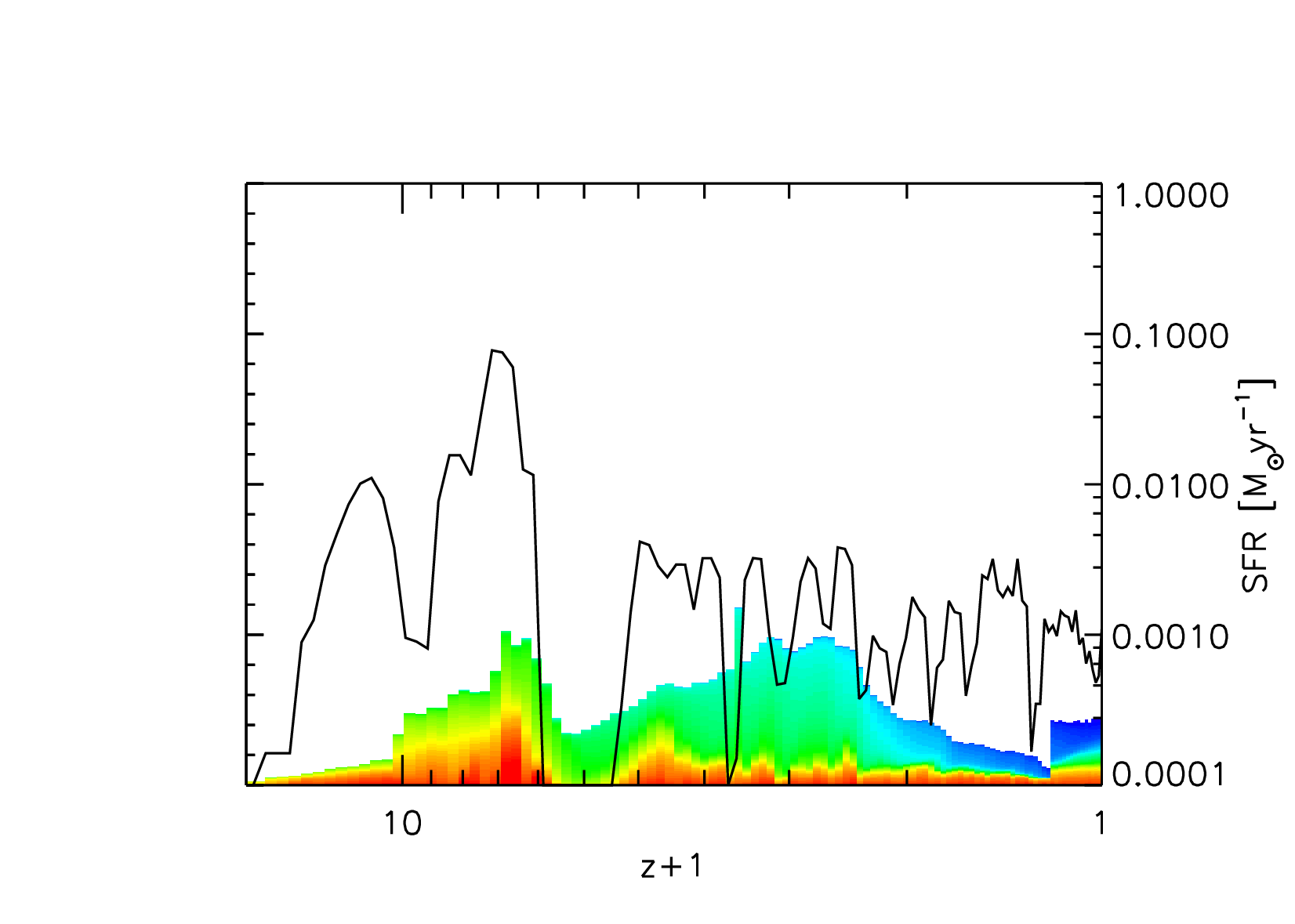} 
      \end{tabular}
  \end{center}

  \caption{Evolution of total gas mass and star formation rate as a
    function of redshift. The top row shows, from left to right,
    haloes 1-3, while the bottom row shows haloes 4-6. The coloured
    area indicates the total amount of gas in each halo in units of
    $\Ms$, corresponding to the scale on the left. The colour coding
    indicates the differential amount of gas at a given mass density
    in units of g cm$^{-3}$, as denoted by the colour bars
    above. Overplotted in black is the star formation rate as a
    function of redshift, in units of $\Ms$~yr$^{-1}$, corresponding
    to the scale on the right. In each halo, star formation is tightly
    coupled to the amount of dense gas, and occurs in bursts, often
    separated by several 100 Myrs, and associated both with supernova
    feedback and with mergers. Halo 5 assembles earlier than the five
    others, and its significantly higher mass at early times leads to
    a more prolonged starburst in the galaxy contained within it. All
    galaxies are star-forming at $z=0$.  \label{fig:evolution}}

\end{figure*}

Supernova energy is shared equally between the hot and cold
phases. Cold particles which receive supernova feedback accumulate
energy until their thermodynamic properties are comparable to those of
their local hot neighbours. At this point, the energy is released and
the particles are promoted to the hot phase. We supplement
\cite{Scannapieco-2006} by including a seeding mechanism that defines
reasonable properties for the local hot phase, even if no neighbouring
particles are considered hot at the time. A cold gas particle which
has received sufficient supernova energy to raise its thermodynamic
properties to this level can thus be promoted, even if it currently
has no hot neighbours. This ensures that the distribution of supernova
feedback is not delayed at the earliest stages of star formation, when
the entire interstellar medium can be in a cold and dense
configuration (see Figure~\ref{fig:evolution}). The seeding mechanism
does not create heat artificially and conserves energy. We have
checked that the amount of gas required to seed the hot phase is
small, and that the ensuing evolution of the two phases is consistent.

\section{Galaxy Formation}\label{sec:evolution}

The different merger histories of the dark matter haloes described in
Section~\ref{sec:ICs} are reflected in their gas accretion histories,
and in the evolution of the galaxies that form within
them. Section~\ref{sec:gx-evolution} describes the co-evolution of the
halo and its galaxy, while Section~\ref{sec:properties} discusses the
properties of the final objects.

\subsection{Galaxy Evolution}\label{sec:gx-evolution}
The halo assembly histories vary significantly, both among our
selected sample and among the total population of $10^{10}\Ms$
haloes. As shown in Figure~\ref{fig:haloes}, haloes 3 and 6 had recent
major mergers at $z=0.32$ and $z=0.21$, respectively, whereas haloes 1
and 2 experienced their last major mergers around $z=1$. Haloes 4 and
5 have not undergone any major mergers since before $z=2$. Halo~5 is
also significantly more massive compared to the other haloes at high
redshift.

The six panels in Figure~\ref{fig:evolution} show the gas mass bound
to each of the six haloes as a function of redshift. While the total
coloured area indicates the total amount of gas in each halo, the star
formation rate in the main progenitor, overplotted in black, depends
on the presence of cold and dense gas, shown in red and orange
colours. We find that two different mechanisms lead to a burstiness of
star formation in our simulations. On timescales of hundreds of Myrs,
self-regulation of star formation and supernova feedback lead to
periodic variations in the gas density, and periodic star formation
behaviour. This confirms the earlier results of \cite{Pelupessy-2004},
\cite{Stinson-2007}, \cite{Mashchenko-2008}, \cite{Valcke-2008} and
\cite{Revaz-2009}. In addition, gas-rich mergers can induce
starbursts. These bursts are irregular and can be separated by several
Gyrs. For example, the star formation episodes in Halo~4 beginning at
$z=8$, $4.3$ and $2.1$ are preceded by mergers at $z=8.5$, $4.7$ and
$2.4$ with haloes of $2\times10^8$, $3\times10^8$ and $10^9\Ms$,
respectively, which bring in fresh gas. Observations of periodic
bursts lasting several hundred million years have been reported for
three dwarf galaxies by \cite{McQuinn-2009}, while a number of dwarf
galaxies (Leo I \citep{Dolphin-2002}, Leo A \citep{Cole-2007}, IC 10
\citep{Cole-2010}, IC 1613 \citep{Skillman-2003}, DDO 210
\citep{McConnachie-2006} and Carina \citep{Koch-2006}) show extended,
quiescent periods between star formation epochs. \cite{Cole-2010}
suggest mergers and gas accretion as triggers for star formation, but
note that individual bursts and mergers can no longer be linked
observationally after several Gyrs.

The star formation history of each individual galaxy in our
simulations reflects a combination of internal self-regulation via
supernova feedback, and the supply of fresh gas via accretion and
mergers. These two effects largely determine the variance in stellar
mass between the haloes in our simulations; while differences in
merger histories increase the variance, self-regulation via feedback
decreases it. In our sample of six haloes of equal final mass, the
galaxy stellar masses vary by about a factor of two.

Gas-rich mergers after $z=6$ imply that the progenitors did not lose
all their gas due to the UV background. In our simulations, such
mergers occur with haloes that would reach masses above $\sim 10^9\Ms$
by $z=0$. \cite{Sawala-2010} showed that at this mass, a combination
of UV and supernova feedback removes gas efficiently, while UV
radiation alone is not always sufficient. Observations of local group
dwarf spheroidals \citep[e.g.][]{Monelli-2010} also suggest that
reionisation had at most a minor effect on these galaxies.

Some major mergers also contribute stars. The fraction of final
stellar mass formed outside of the main progenitor ranges from
$\sim5\%$ in Halo 5, accreted at $z=5.2$, to close to $40\%$ for Halo
1, resulting from two major mergers at $z=3.4$ and $z=0.8$. Haloes 2
and 6 both accrete $\sim25\%$ in mergers at $z=0.4$ and 1,
respectively, while haloes 3 and 4 accrete $\sim7\%$ in mergers at
$z=0.5$ and $z=2.4$. Since haloes 3, 4 and 5 follow a more typical
assembly history, we expect the typical fraction of stars formed
outside the main progenitor in dwarf galaxies of M$_\star\sim10^8\Ms$
to be $<10\%$, albeit with possible exceptions.

\begin{table} 
\caption{Overview of numerical simulation results\label{table:results}}
\setlength{\tabcolsep}{2pt}
\begin{tabularx}{\columnwidth}{lccccccr}
  \hline 
   Halo& M$_\star$   & M$_\mathrm{g}$ & M$_\mathrm{DM}$ & r$_{1/2}$ & $\sigma_\star$ &L$_\star$ & Z$_\mathrm{MM}$ \\
     &\scriptsize{[$10^7\Ms$]}& \scriptsize{[$10^7\Ms$]} &\scriptsize{[$10^9\Ms$]}  &\scriptsize{[kpc]} 
     &\scriptsize{[km s$^{-1}$]} & \scriptsize{[km s$^{-1}$kpc]} &\\
  \hline  \vspace{-.05in}\\
  1 & $7.81$ & $2.81$ & $9.41$ & 0.87 & $21$ & $1.4$ & $0.77$ \\     
  2 & $5.25$ & $6.23$ & $8.34$ & 0.39 & $17$ & $0.7$ & $0.96$ \\     
  3 & $4.94$ & $19.4$ & $8.80$ & 0.28 & $15$ & $1.1$ & $2.65$ \\     
  4 & $8.14$ & $4.17$ & $8.98$ & 0.36 & $19$ & $0.4$ & $2.37$ \\     
  5 & $10.2$ & $7.11$ & $9.94$ & 0.74 & $33$ & $19$  & $5.17$ \\     
  6 & $6.17$ & $5.07$ & $8.54$ & 0.68 & $18$ & $0.3$ & $0.38$ \\     
  \hline \vspace{.1in}
\end{tabularx}

\small{Notes: Col.~2: Stellar mass, Col.~3: Gas mass, Col.~4: Dark
  matter mass (all measured inside r$_{200}$ at $z=0$), Col.~5:
  Stellar half-mass radius, Col.~6: De-projected 1-D RMS stellar
  velocity dispersion, Col.~7: Specific stellar angular momentum
  $\mathrm{L} = |\mathbf{r} \times \mathbf{v}|$ ($\sigma_\star$ and
  L$_\star$ measured within 3~kpc), Col.~8: Redshift of last major
  merger (progenitor mass ratio $<~3:1$).}
\end{table}

In general, we find that the mean metallicity evolves with age,
indicating the recycling of enriched gas in subsequent generations of
stars. At each stellar age, we also find a spread in metallicities,
which indicates the incorporation of fresh material. However, we note
that due to a lack of diffusive metal mixing, the metallicity spread
in our simulated stellar populations can be as high as 3 dex, which is
larger than observed.

\subsection{Galaxy Properties}\label{sec:properties}
Several properties of the six simulations are listed in
Table~\ref{table:results}. They appear to be in broad agreement with
the previous studies discussed in Section~\ref{sec:previous}. The
final baryon fraction of the haloes are between 1.1 and 2.7 \% of the
total matter. The final stellar masses of the six galaxies fall
between $4.9 \times 10^7$ and $1.0\times10^8 \Ms$, which corresponds
to stellar mass to total mass ratios in the range of
$\sim~5\times10^{-3}-10^{-2}$. All galaxies are star-forming at
$z=0$.

The final gas masses vary from $2.8\times10^7$ to
$1.9\times10^8\Ms$. Observed dwarf galaxies of this stellar mass
typically have a substantial HI content
\citep[e.g.][]{Staveley-Smith-1992, Geha-2006, deBlok-2008}. In our
simulations, we cannot directly measure the amount of HI gas. Defining
the cold gas as the total amount of gas at temperatures below the peak
of the cooling curve and correcting for the contribution of helium and
metals, we derive approximate upper limits for the HI masses between
$10^7$ and $1.2\times10^8\Ms$. We adopt the notation of
\cite{Geha-2006} in defining the HI mass fraction as
$f_\mathrm{HI}=\mathrm{M}_\mathrm{HI}/(\mathrm{M}_\mathrm{HI}+\mathrm{M}_\star)$,
noting that He, H$_2$, hot gas and metals are neglected in the
denominator. We derive upper limits for $f_\mathrm{HI}$ between $11\%$
and $71\%$ in our six galaxies with a median of $35\%$. In a sample of
101 flux-selected SDSS dwarf galaxies of similar stellar mass,
\citeauthor{Geha-2006} find a higher mean HI fraction of
60$\%$. However, individual galaxies show a large scatter in
$f_\mathrm{HI}$, with several as high as $95\%$, and others with upper
limits below $10\%$.

The range of stellar half-mass radii in our six simulations is 0.28 to
0.87~kpc. This is comparable to the typical values obtained by \cite{
  Geha-2006} for half-light radii (r-band), even though the observed
sample also contains a handful with half-light radii greater than
1.5~kpc. We note that only Halo~5, which has the most quiescent
assembly history, contains a galaxy with a rotationally supported
stellar disk, as reflected by the specific angular momentum
L$_\star$. The other five haloes have more ellipsoidal morphologies
and very little rotation. This is in contrast to the result of
\cite{Governato-2010}, who report pristine disk galaxies in both of
their simulations, albeit in haloes of 2.0 and $3.5\times
10^{10}\Ms$. Observations also suggest that isolated dwarf galaxies of
this stellar mass are more frequently disk-like
\citep[e.g.][]{Hunter-2006, Geha-2006}. The latter find that $30\%$ of
edge-on dwarf galaxies show coherent rotation profiles, a number which
drops to $18\%$ when all axis-ratios are included, but this should be
considered as a lower limit. In a volume-limited sample,
\cite{Sanchez-2010} report that less than $30\%$ of galaxies with
stellar masses of $10^8\Ms$ have apparent axis ratios below $0.5$, and
less than $5\%$ below $0.3$. They attribute this flattening to stellar
feedback. Compared to these observations, our simulated sample of six
haloes is too small to assess the statistical significance of a 1 in 6
result. It should also be noted that the stellar half-mass radii of
our galaxies are only resolved with a few softening
lengths. Therefore, results on details of the galactic structure from
our simulations are inconclusive.

\section{Stellar Mass~--~Halo Mass Relation} \label{sec:sams}
In Section \ref{sec:ICs}, we explained how we constructed our initial
conditions from the parent simulation. Its large volume and high
dynamic range allows us to demonstrate that we have resimulated a
representative sample of haloes. We can therefore derive implications
for the global population of galaxies that form in similar mass
haloes, and compare with expectations from matching the observed
stellar mass function to the abundance of haloes in a $\Lambda$CDM
universe.

\begin{figure}
  \vspace{-.15in}
  \hspace{-.2in} \includegraphics[scale = .5]{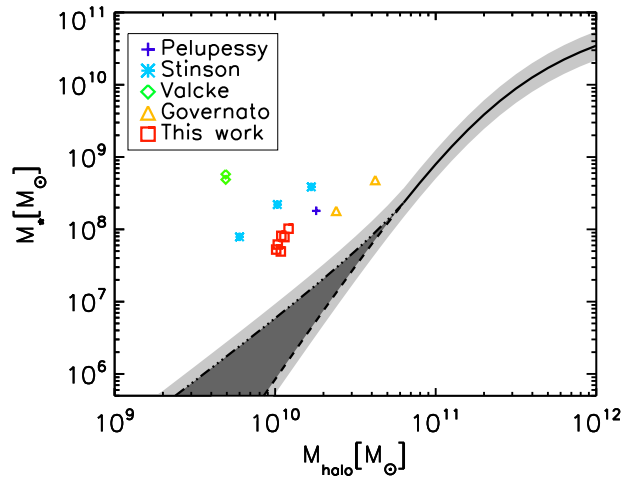}
  \caption{The stellar mass~--~halo mass relation, derived by Guo et
    al. (2010), compared to the results of several numerical
    simulations. The solid black line denotes the range constrained by
    the SDSS DR-7 data, while the dashed line is an extrapolation to
    lower masses, with a faint-end slope of $-1.15$ for the stellar
    mass function. The dark grey area shows the influence of a steeper
    faint-end slope up to $-1.58$ (dash-dotted line), while the light
    grey area represents the maximally allowed dispersion of 0.2 dex
    in M$_\star$ for a given halo mass. The coloured symbols denote
    the results of hydrodynamical simulations, as listed in
    Table~\ref{table:other} and Table~\ref{table:results}, excluding
    simulations that did not evolve to $z=0$ or assume a baryon
    fraction of 1\% {\it ab initio}. The red squares indicate our own
    six simulations. For consistency, we use the halo masses of the
    pure dark matter simulations for our own simulations, and apply a
    correction of $20\%$ to all other haloes, where this information
    is not available. All hydrodynamical simulations overpredict the
    stellar mass with respect to the observed relation by more than an
    order of magnitude.\label{M/L-Qi}}
\end{figure}

Assuming a monotonic relationship between stellar mass and maximum
halo mass, \cite{Guo-2010} have compared the abundance of
haloes/subhaloes in the Millennium and Millennium-II~Simulations to
the observed abundance of galaxies as a function of stellar mass
obtained from the SDSS DR-7 by \cite{Li-2009}. The observational
sample contains over half a million galaxies at low redshift, and
extends down to stellar masses of $10^{8.3}\Ms$ with very small error
bars. The combination of the two very large simulations also leads to
very small errors on the theoretical halo abundance. The derived
stellar mass to halo mass ratio peaks at $M_{halo} = 10^{11.8} \Ms$ at
a star formation efficiency of about $20\%$, and decreases both for
more massive and for less massive haloes (see Figure
\ref{M/L-Qi}). The decrease in efficiency at the high mass end is
generally attributed to AGN feedback \citep[e.g.][]{Croton-2006,
  Bower-2006}, while the decrease for lower mass haloes is assumed to
be due to the increasing efficiency of supernova feedback, and the
effect of the UV background. This general behaviour was noted earlier
from lower precision data by \cite{Navarro-2000}, \cite{Yang-2003},
\cite{Dekel-2003}, \cite{Conroy-2009} and \cite{Moster-2009}.\\

Following \cite{Yang-2003}, \cite{Guo-2010} adopted the following
functional form for the mean stellar mass to halo mass ratio:
$$ \frac{M_\star}{M_{halo}} = c
\left[\left(\frac{M_{halo}}{M_0}\right)^{-\alpha} +
  \left(\frac{M_{halo}}{M_0}\right)^{\beta}\right]^{-\gamma}
$$ They report an accurate fit to the data with parameters $c=0.129$,
$M_0 = 10^{11.4}\Ms$, $\alpha=0.926$, $\beta=0.261$ and
$\gamma=2.440$. At the low mass end, the SDSS DR-7 data of
\citeauthor{Li-2009} extends to stellar masses of $2\times10^8\Ms$
with high accuracy, corresponding to a halo mass of
$10^{10.8}\Ms$. \citeauthor{Guo-2010} extrapolate this relation down
to the lowest halo masses resolved in the MS-II. For halo masses of
$\sim10^{10}\Ms$, this predicts a stellar mass of $\sim8\times10^5
\Ms$.

From a similar analysis based on SDSS~DR-3, which extends to stellar
masses of $3.2\times10^8\Ms$, \cite{Moster-2009} also derive a stellar
mass~-- halo mass relation, in good agreement with \cite{Guo-2010} at
the high mass end. They predict haloes of $10^{10}\Ms$ to host
galaxies with stellar masses of $\sim5.7\times10^6\Ms$, significantly
higher than found by \citeauthor{Guo-2010}, but still an order of
magnitude lower than our hydrodynamical simulations
predict. \citeauthor{Moster-2009} note, however, that such haloes are
at best marginally resolved at their mass resolution ($2.8\times 10^8
\Ms$), prohibiting a self-consistent treatment of
subhaloes. Consequently, they only include haloes above
$1.6\times10^{10}\Ms$ in the conditional mass
function. \citeauthor{Moster-2009} also apply their analysis to an
analytic Sheth-Tormen mass function obtained by \cite{Vale-2006}. In
this non-parametric model, a halo mass of $10^{10}\Ms$ corresponds to
a stellar mass of $1.9\times10^6\Ms$, more similar to the value of
\cite{Guo-2010}.

In Figure~\ref{M/L-Qi}, we plot the stellar mass~-- halo mass relation
of \cite{Guo-2010} for haloes between $10^{9}$ and $10^{12}\Ms$. The
solid section of the line shows the relation in the region directly
derived from SDSS DR-7 data where the uncertainties are very
small. The dashed section denotes an extrapolation to stellar masses
below $10^{8.3}\Ms$, assuming a faint-end slope of $\alpha = -1.15$
for the stellar mass function, as reported by \cite{Li-2009}.  Studies
of the faint-end of the stellar mass function are either limited to
nearby regions or galaxy clusters, or require corrections for
incompleteness and, in the case of photometric redshifts, background
subtraction, which introduce considerable uncertainties
\citep[e.g.][]{Christlein-2009}. As a result, different values for
$\alpha$ in the range of $-1.1$ to $-1.6$ are found in the recent
literature \citep[e.g.][]{Trentham-2005, Blanton-2005, Carrasco-2006,
  Baldry-2008}. The dark grey area in Figure~\ref{M/L-Qi} shows the
effect of a steepening of the faint-end slope up to $\alpha = -1.58$,
the value reported by \cite{Baldry-2008}. While this has a strong
effect on the lowest mass haloes, we note that it cannot account for
the discrepancy we find in haloes of $10^{10} \Ms$. In order to fit
the constraints of SDSS DR-7, the maximal dispersion at fixed halo
mass is 0.2 dex in M$_\star$, indicated by the light-grey area. We
overplot the results of our six simulations as red squares and add
other $z=0$ predictions from the studies listed in
Table~\ref{table:other}, correcting all halo masses for baryonic
effects, as described below. It is apparent that all these
hydrodynamical simulations overproduce stellar mass for their
respective halo mass by at least an order of magnitude.

In Table~\ref{table:compare}, we compare the properties of our six
simulations to the abundance matching predictions. We note that, due
to the outflow of baryons, the total mass of our six haloes is almost
a factor of $1-\Omega_b/\Omega_m$ smaller than the masses of the
corresponding haloes from the pure dark matter simulation. This effect
is expected at such low star formation efficiency. For consistency
with \cite{Guo-2010}, we therefore use the (higher) peak masses of the
pure dark matter simulations in deriving the stellar mass predicted
for each of our haloes by the abundance matching argument. For all
galaxies listed in Table~\ref{table:other} whose peak halo mass cannot
be defined or is not given, we increase the halo mass in
Figure~\ref{M/L-Qi} by $\Omega_b/\Omega_m\sim20\%$, the maximally
expected correction.

Comparing the results of our simulations to the predictions, we find
that the hydrodynamical simulations overproduce stellar mass by a
median factor of $\sim 50$. Alternatively, abundance matching predicts
that galaxies of $10^{7.9}\Ms$, the median stellar mass produced in
our hydrodynamical simulations, should reside in haloes with typical
masses of $\sim 4.5\times10^{10}\Ms$, rather than $10^{10}\Ms$. If
$10^{10}\Ms$ haloes really hosted galaxies with $M_\star=10^{7.9}\Ms$,
a $\Lambda$CDM universe would overpredict their abundance by a factor
of $\sim 4$.

\begin{table}
\caption{Comparison of stellar mass~-- halo mass
  ratios\label{table:compare}}

\begin{tabularx}{\columnwidth}{lrrrr}
  \hline
  Halo & M$_\star$ & M$_\mathrm{tot}$ & M$_\mathrm{max}$ & M$_\star$(SMF) \\ 
  &\scriptsize{$[10^7\Ms]$}& \scriptsize{$[10^9\Ms]$}  &\scriptsize{$[10^9\Ms]$}& \scriptsize{$[10^7\Ms]$}  \\
  \hline  \vspace{-.05in}\\
  1 & $7.81$ & $9.52$ & $12.1$ & $0.154$ \\ 

  2 & $5.25$ & $8.46$ & $11.2$ & $0.1216$ \\

  3 & $4.94$ & $9.04$ & $10.7$ & $0.104$ \\

  4 & $8.14$ & $9.10$ & $11.8$ & $0.143$ \\

  5 & $10.2$ & $10.1$ & $12.5$ & $0.171$ \\

  6 & $6.16$ & $8.65$ & $10.7$ & $0.104$ \\
\hline \vspace{.1in}
\end{tabularx}

\small{Notes: Col.~2: Stellar mass obtained from simulation, Col.~3:
  Combined mass (m$_{200}$) of stars, gas \& dark matter of the halo
  in the hydrodynamical simulation, Col.~4: Peak halo mass (m$_{200}$)
  in the pure dark matter simulation, Col.~5: Stellar mass
  corresponding to M$_\mathrm{max}$ from the abundance matching.}
\end{table}

This discrepancy is too large to be attributed solely to
incompleteness in the observed stellar mass
function. \cite{Baldry-2008} have used the stellar mass~-- surface
brightness relation of SDSS galaxies in order to estimate the
completeness at the faint end. Based on this analysis, \cite{Li-2009}
estimate the completeness at $10^{8.3}\Ms$ to be well above
$70\%$. Following \citeauthor{Baldry-2008}, the uncertainty in the
number of $10^{7.9}\Ms$ galaxies is much smaller than the discrepancy
we report.

The difference is also unlikely to be attributable to numerical errors
in our hydrodynamical simulations, or to the specific parametrisation
of star formation and feedback in our model. From
Table~\ref{table:other}, it is clear that all other current
hydrodynamical models, while succeeding in reproducing many of the
observed features of individual dwarf galaxies, also predict similar
or higher galaxy mass to halo mass ratios. They thus also fail to
reproduce the low star formation efficiencies required to explain the
observed abundances of dwarf galaxies in a $\Lambda$CDM universe.

Haloes of $10^{10}\Ms$ are well resolved in the Millennium-II
Simulation, and the number of such haloes in a volume of
$137^3$~Mpc$^3$ is clearly large enough for statistical uncertainties
to be small. While the assumed cosmological parameters of
$\Omega_\Lambda = 0.75$, $\Omega_\mathrm{m} = 0.25$ and $\sigma_8 =
0.9$ are only marginally consistent with the five-year WMAP data
\citep{Komatsu-2009}, this has a negligible effect on the mass
function. It is unlikely that the number density of $10^{10} \Ms$
haloes formed in a $\Lambda$CDM cosmology is significantly
overestimated in our parent simulation. A lower value of $\sigma_8$
would slightly {\it increase} the abundance of haloes of this mass,
but as \cite{Yang-2003} have shown, the abundance matching result,
which depends on the cumulative abundance of more massive haloes, is
almost unchanged at $10^{10} \Ms$ in $\Lambda$CDM.

In warm dark matter (WDM) models, structure is erased below a
characteristic free-streaming length that depends on the assumed
properties of WDM particles. \cite{Zavala-2009} have compared the halo
mass functions in high resolution dark matter simulations of
$\Lambda$CDM and $\Lambda$WDM, assuming $m_\mathrm{WDM} = 1$~keV, and
find that the present-day abundance of haloes of $10^{10}\Ms$
decreases by a factor of $\sim 3$. These simulations truncate the
power spectrum of the initial conditions in the WDM case but neglect
thermal velocities, which could increase the effect. However, recent
combined analysis of structures observed in the Lyman-$\alpha$ forest
of SDSS and HIRES by \cite{Viel-2008} suggest a lower limit of $\sim
4$~keV for thermal relics in a pure $\Lambda$WDM cosmology, which
would allow only a much smaller deviation from CDM on these
scales. While a WDM model could thus perhaps account for the reported
discrepancy between simulations and star formation efficiencies
inferred from abundance matching, the required WDM particle mass
appears disfavoured by observation. A WDM solution would also
significantly alter the internal structure of dwarf haloes, and
simultaneously reduce the abundance of lower mass objects.

In principle, one could turn to direct measurements of halo masses for
individual galaxies on the relevant scales, to elucidate whether the
halo mass function of the $\Lambda$CDM model, and the inferred stellar
mass~-- halo mass relationship, are correct. Direct mass estimates,
through gravitational lensing \citep[e.g.][]{Mandelbaum-2006} are only
available for haloes with masses above $\sim10^{11.8}\Ms$, however,
where they agree with the CDM predictions and the relationship of
\citeauthor{Guo-2010}. For dwarf galaxies, one has to rely on HI
rotation curves \citep[e.g.][]{deBlok-2008}, which do not give
reliable estimates for {\it total} halo masses.

As demonstrated in Section~\ref{sec:ICs}, we have been careful to
resimulate a representative sample of haloes, and to exclude any
systematic bias. Considering the limited variance in stellar mass
among the six haloes, statistical fluctuations are an unlikely source
for the discrepancy.

It is worth noting that semi-analytical models of galaxy formation
\citep[e.g.][]{Kauffmann-1993, Cole-2000} attempt to reproduce the
observed faint-end slope of the stellar mass function in a
$\Lambda$CDM universe by assuming highly efficient supernova feedback
in small haloes \citep[e.g.][]{Benson-2003, Khochfar-2007}. We have
applied the semi-analytic model of \citet[b]{Guo-SAM}, which
reproduces the \cite{Li-2009} stellar mass function, to the merger
trees of our six resimulated haloes, and also to a randomly selected
sample of similar mass haloes from the Millennium-II Simulation. We
find no difference in the predicted stellar mass between the selected
haloes and the random sample, but stellar masses that are roughly two
orders of magnitude smaller than in our hydrodynamical
simulations. Independent of the cause of the discrepancy between the
hydrodynamical simulations and the observed stellar mass function,
there is thus a divergence between current hydrodynamical and
semi-analytical models for dwarf galaxies. The semi-analytical models
are tuned to produce the correct galaxy abundance, whereas
hydrodynamical models aim at reproducing the physical processes and
the structure of individual galaxies. Clearly, these two aspects
cannot be treated separately, if we are to converge to a consistent
picture of galaxy formation.

\section{Summary}\label{sec:summary}
We have performed high-resolution hydrodynamical simulations of six
$\sim 10^{10}\Ms$ haloes, extracted from a large, cosmological parent
simulation. We find that differences in merger histories lead to the
formation of dwarf galaxies with different star formation histories
and final stellar masses between $4.9\times 10^7$ and $10^8\Ms$. These
stellar masses agree with previous simulations of similar mass haloes,
and the structure of our simulated galaxies resembles that of observed
galaxies of similar stellar mass, to the extent which we can resolve
structure in our simulations.

However, all these simulations imply an efficiency of conversion of
baryons into stars which is at least an order of magnitude larger than
that which is required to explain the observed abundance of dwarf
galaxies in a $\Lambda$CDM universe. While current hydrodynamical
simulations, including our own, are consistent with almost arbitrarily
high mass-to-light ratios for the faintest galaxies in haloes of
$10^9\Ms$ or less, they thus appear to be inconsistent with the
mass-to-light ratios of larger dwarf galaxies, even when a moderately
steep faint-end slope of the stellar mass function is assumed. The
current recipes for mechanisms such as UV heating and supernova
feedback appear sufficient to remove the ``Missing Satellites
Problem'' for the smallest satellites. However, isolated dwarf
galaxies with stellar masses of $10^{8}\Ms$ are still substantially
overproduced in current hydrodynamical simulations, even when these
mechanisms are included.

Our results suggest three possible explanations: The current
observational count of dwarf galaxies could be incomplete,
underestimating the true number density of $10^8\Ms$ galaxies by a
factor of four or more. In that case, the hydrodynamical simulations
could be correct, but the semi-analytical models that produce low
abundances of dwarf galaxies have been tuned to incorrect data.

If the count of dwarf galaxies is almost complete at $10^8\Ms$, these
galaxies must, in a $\Lambda$CDM universe, be residing in haloes
significantly more massive than $10^{10}\Ms$, and all current
hydrodynamical simulations overpredict the efficiency of star
formation by more than a factor of ten. This could be an indication of
numerical problems, or, more likely, of incorrect or incomplete
assumptions about the relevant astrophysics. Several possible
  mechanisms may contribute to a star formation efficiency in current
  simulations that is too high compared to real galaxies: 
\begin{itemize}
\item Supernova feedback may be more efficient in ejecting gas from
  dwarf galaxies than current hydrodynamical simulations predict. For
  example, \citet[b]{Guo-SAM} showed that the observed stellar mass
  function can be reproduced in semi-analytic models by assuming very
  strong mass-loading of winds in low mass haloes.
\item The full effect of reionisation on the IGM may not be captured
  in current models. As a result, cooling times may be underestimated,
  and the fraction of gas-rich mergers overestimated. Local sources of
  extreme UV and soft X-ray radiation may also ionise the interstellar
  medium, inducing another self-regulation mechanism for star
  formation \citep{Ricotti-2002, Cantalupo-2010}.
\item Low dust content may lead to less efficient cloud formation and
  shielding at low metallicities. The transformation rate of cold gas
  into stars, currently assumed to be universal, may therefore be
  overestimated in dwarf galaxy simulations \citep{Gnedin-2008}.
\item Processes such as magnetic fields, cosmic rays and the feedback
  from population-III stars are not included in any of the current
  models, and may further reduce the star formation efficiency.
\end{itemize}

\noindent Any revised model would however still have to reproduce
features of individual galaxies consistent with observations. We also
note that a model which substantially decreases the number of
$10^8\Ms$ galaxies would imply that the halo masses of fainter dwarf
galaxies would need to be revised upwards, as some of these would now
be required to live in $10^{10}\Ms$ haloes.

If the observed stellar mass function is complete, and the
hydrodynamical simulations correctly capture the relevant physics of
galaxy formation, the Millennium-II Simulation (and similar
$\Lambda$CDM simulations) overpredict the number of $10^{10}\Ms$ dark
matter haloes. This would seem to require the underlying physical
assumptions of the $\Lambda$CDM model to be revised. Warm Dark Matter
may offer a possibility, but only for particle masses of $\sim1$~keV,
below the limit apparently implied by recent Lyman-$\alpha$
observations.

Of the three proposed scenarios, it appears that missing astrophysical
effects in the simulations are the most likely cause of the
discrepancy, and the most promising target in search of its
resolution. While the three scenarios differ in nature, none is
without significant implications for galaxy formation, which will have
to be addressed in the future.

\section*{Acknowledgements}
We would like to thank Volker Springel for the numerical methods that
made this work possible, and Cheng~Li and Mike Boylan-Kolchin for the
many helpful discussions. We also thank our anonymous referee for the
helpful comments and suggestions that have improved our
manuscript. All simulations were carried out at the computing centre
of the Max-Planck Society in Garching. \bibliographystyle{mn2e}

\bibliography{manuscript}

\label{lastpage}

\end{document}